\documentclass{aa}

\usepackage{natbib}
\usepackage[english,french]{babel}

\usepackage[latin1]{inputenc}
\usepackage[T1]{fontenc}

\usepackage{url}
\usepackage{xspace}
\usepackage{graphicx}
\usepackage{txfonts}

\graphicspath{{fig/}}

\providecommand{\unit}[1]{\ensuremath{\:\mathrm{#1}}}
\providecommand{\textind}[1]{{\ensuremath{\textrm{\scriptsize{#1}}}}}

\newcommand{\tise}[1]{<#1>\xspace}

\newcommand{\tomodifylater}[1]{[\textsl{\textbf{#1}}]}

\newcommand{\changed}[1]{\textbf{#1}}
\renewcommand{\tomodifylater}[1]{}
\renewcommand{\changed}[1]{#1}

\newtheorem{defin}{Definition}{\bf}{\rm}
\newtheorem{variant}{Variant}[defin]{\bf}{\rm}
\makeatletter
\renewcommand{\thevariant}{\@arabic\c@defin.\@arabic\c@variant}
\makeatother

\def\eg{{\it e.g.}\ }

\def\ie{{\it i.e.}\ }

\newcommand{\ud}{\ensuremath\,\mathrm{d}}

\newcommand{\epsthr}{\ensuremath\epsilon_\textind{thr}}

\begin{document}
\selectlanguage{english}

\title{Influence of the definition of dissipative events on their statistics}
\author{E. Buchlin\inst{1,2}
  \and S. Galtier\inst{1}
  \and M. Velli\inst{2,3}
}

\offprints{E. Buchlin, \protect\url{eric.buchlin@ias.fr}}

\institute{
  Institut d'Astrophysique Spatiale, CNRS -- Universit\'e Paris-Sud,
  B\^at.~121, 91405 Orsay Cedex, France
  \and
  Dipartimento di Astronomia e Scienza dello Spazio, Universit\`a
  di Firenze, 50125 Firenze, Italy
  \and
  Istituto Nazionale Fisica della Materia, Sezione A, Universit\`a di
  Pisa, 56100 Pisa, Italy
}

\date{Received\,:  / Revised date\,:}

\abstract{ A convenient and widely used way to study the turbulent plasma in
  the solar corona is to do statistics of properties of events (or
  structures), associated with flares, that can be found in observations or
  in numerical simulations.  Numerous papers have followed such a
  methodology, using different definitions of an event, but the reasons
  behind the choice of a particular definition (and not another one) is very
  rarely discussed.  We give here a comprehensive set of possible event
  definitions starting from a one-dimensional data set such as a time-series
  of energy dissipation.  Each definition is then applied to a time-series
  of energy dissipation issued from simulations of a shell-model of
  magnetohydrodynamic turbulence as defined in \citet{giu98}, or from a new
  model of coupled shell-models designed to represent a magnetic loop in the
  solar corona. We obtain distributions of the peak dissipation power, total
  energy, duration and waiting-time associated to each definition. These
  distributions are then investigated and compared, and the influence of the
  definition of an event on statistics is discussed.  In particular,
  power-law distributions are more likely to appear when using a threshold.
  The sensitivity of the distributions to the definition of an event seems
  also to be \changed{weaker} for events found in a highly intermittent time
  series.  Some implications on statistical results obtained from
  observations are discussed.


}

\maketitle

\section{Introduction}

The mechanism heating the solar corona to millions of degrees remains an
open problem, but it is generally understood that a great part of the energy
dissipation must occur at scales that are smaller than the structures that
can be resolved by observations ($\approx 100\unit{km}$), as small as
$10$--$100\unit{m}$ (the Kolmogorov turbulent cascade dissipation scale).
One of the most successful approaches to reach this four-order-of-magnitude
wide gap is to assume that the statistics obtained at observable scales are
still valid at smallest scales.  The properties of the global system, from
observable to non-observable scales, can then be investigated.  This is for
example the idea underlying Hudson's (1991)\nocite{hud91} critical power-law
slope of $-2$ for the distribution of flare energies.

The measurement of the power-law slope for the lowest energy flares has
indeed been a major goal of coronal physics in the last decade.
\citet{asc00} has summarized the distributions of event energies that were
obtained at wavelengths from X-rays to ultra-violet (UV), and for event
energies covering a range of eight orders of magnitude from
$10^{17}\unit{J}$ (``nanoflares'') to $10^{25}\unit{J}$ (``flares''). It
seems --- and it is a statement of \citet{asc00} --- that these
distributions can be matched together to form a unique power-law
distribution of slope $\approx -1.8$.

However, for the smallest events, mainly observed in UV by filter imaging
instruments, some observations seem not to mutually agree, and they seem not
to fit in the global distribution. A possible explanation for this is given
by \citet{asccha02}: the energy of an event could be wrongly deduced from
the observable quantities (like intensity in some spectral lines), leading
to a systematic error in the distribution of event energies. Another
explanation could be that all authors do not agree on what they mean by
``event'', \ie the fact that inequivalent definitions exist in the
literature.

Indeed, defining an event is likely to be much more difficult for low energy
events than for high energy events. For high-energy events, whose
distributions are in general derived from X-rays and radio observations,
there is little ambiguity on what is an event: events are very rare
($10^{-6}\unit{s^{-1}}$ for the whole solar disk between $10^{23}$ and
$10^{24}\unit{J}$) and well-separated by long low-flux times. On the
contrary, low-energy events can be very close in space and time, making it
difficult to separate them, either because they occur on the same
line-of-sight or because they are smaller than the instrumental resolution
or shorter than the time resolution.  The difficulty is even bigger if we
subscribe to the idea that the corona is in a self-organized critical state
so that small events trigger other events, leading to avalanches as
illustrated by the sandpile paradigm \citep{bak88,luh91}: among all these
events, which ones should be used to do statistics?

We think that this difficulty has been underestimated when statistics
obtained from observations or simulations by different methods have been
compared. In fact, the definition of an event which is used is very rarely
discussed (contrary to the influence of the relationship between the
observable parameters and the physical variables of an event), and is
sometimes even not given precisely.

In this paper we give some definitions that could be used, mainly inspired
from definitions which have already been used in the past. We choose to
restrain ourselves to events defined from a one-dimensional data set, namely
a time series of energy dissipation, so that the definitions can be easily
compared. We then produce statistics of events (histograms of event
energies, durations and waiting times), for different definitions of an
event, and compare them. The time series we use are the data output by a
shell-model of MHD turbulence \citep{giu98}, and by a new model of coupled
shell-models describing Alfvén turbulence in a coronal loop
\citep{buc04soho15}. However, the aim of this paper is not to study
shell-models of MHD, but to see to which extent the definition of events
influences their statistics, even in a simple case of events detected in a
time series of energy dissipation.

\section{Event definitions}

\begin{figure*}
  \begin{minipage}{.33\linewidth}
    \includegraphics[width=\linewidth]{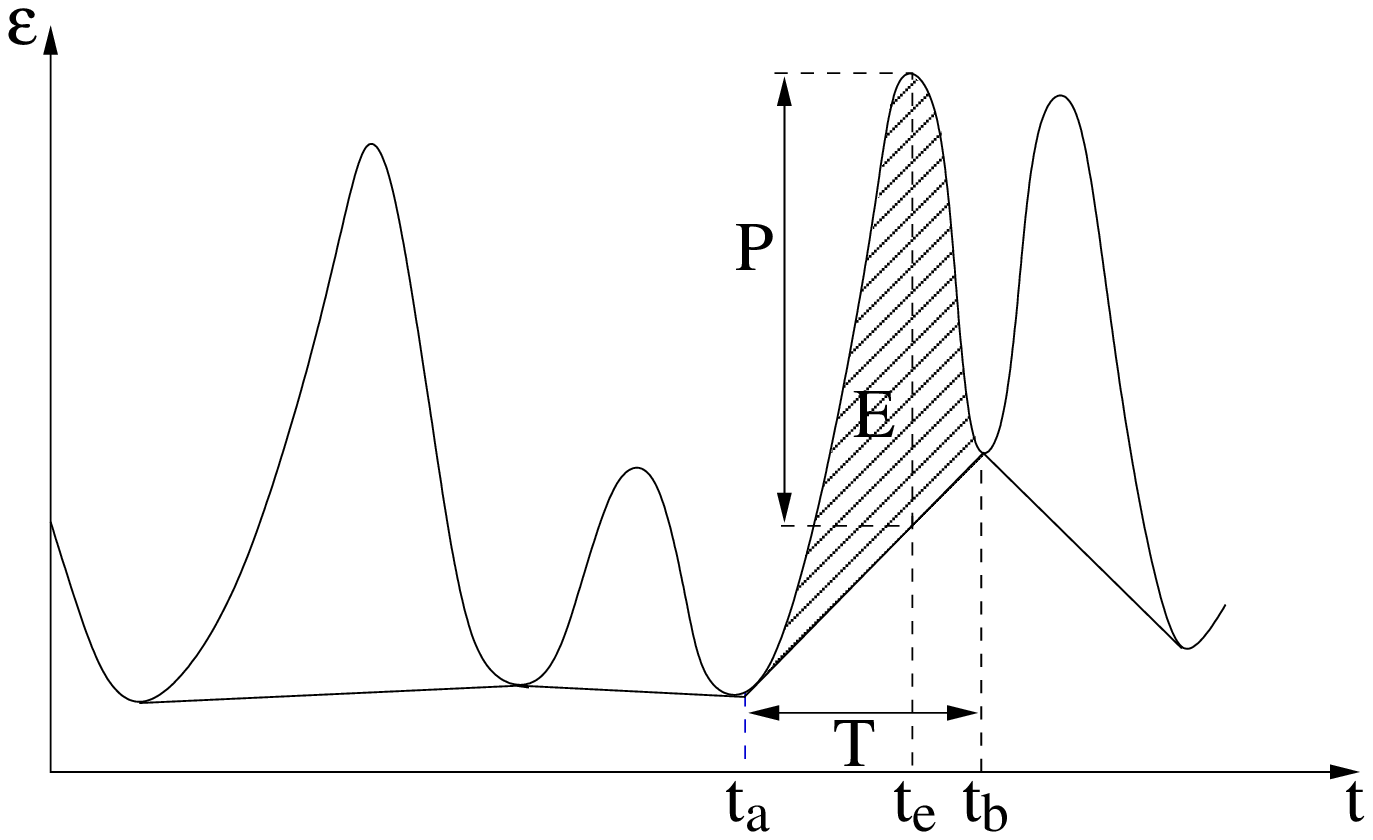}\\
    \includegraphics[width=\linewidth]{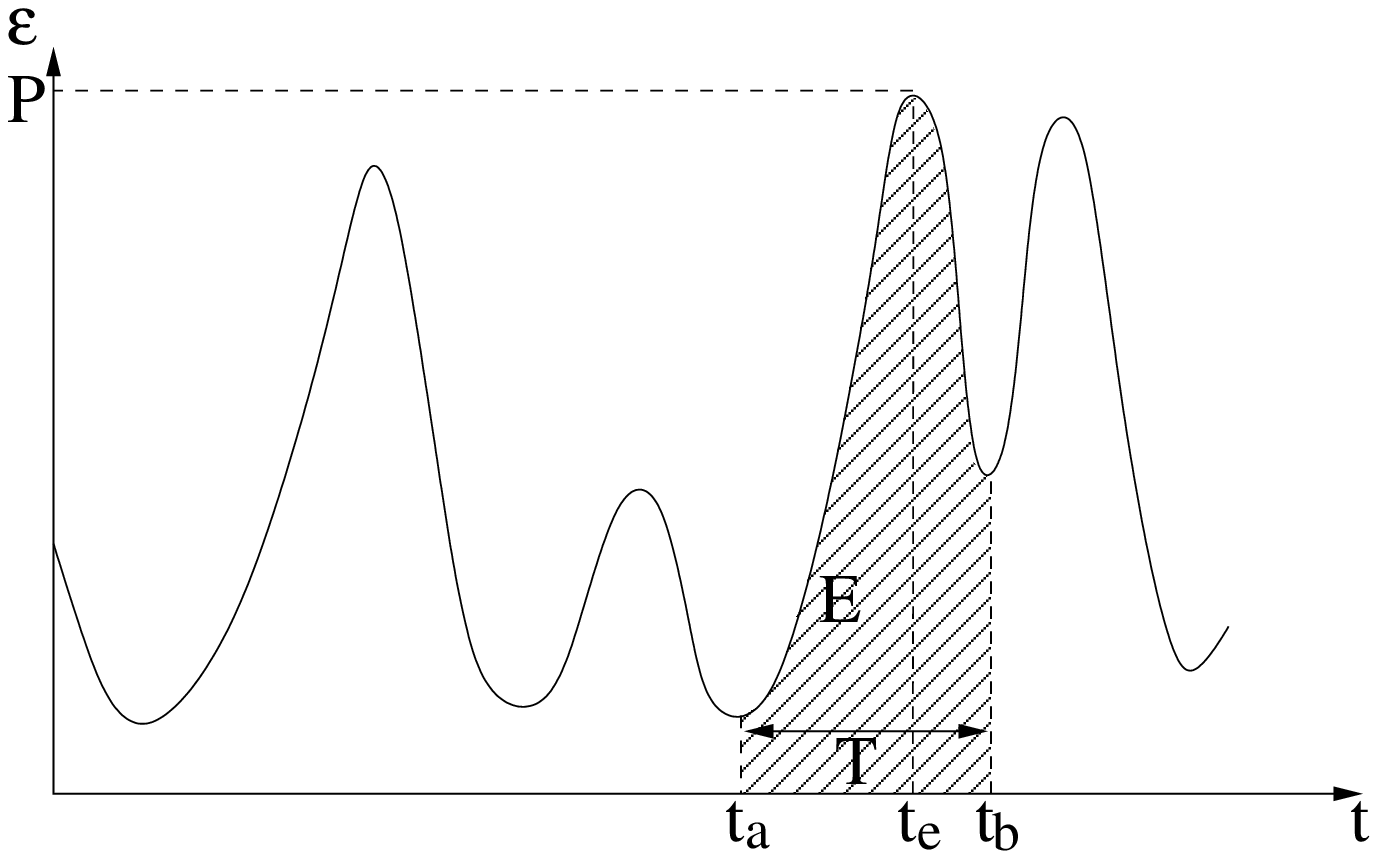}
  \end{minipage}
  \begin{minipage}{.33\linewidth}
    \includegraphics[width=\linewidth]{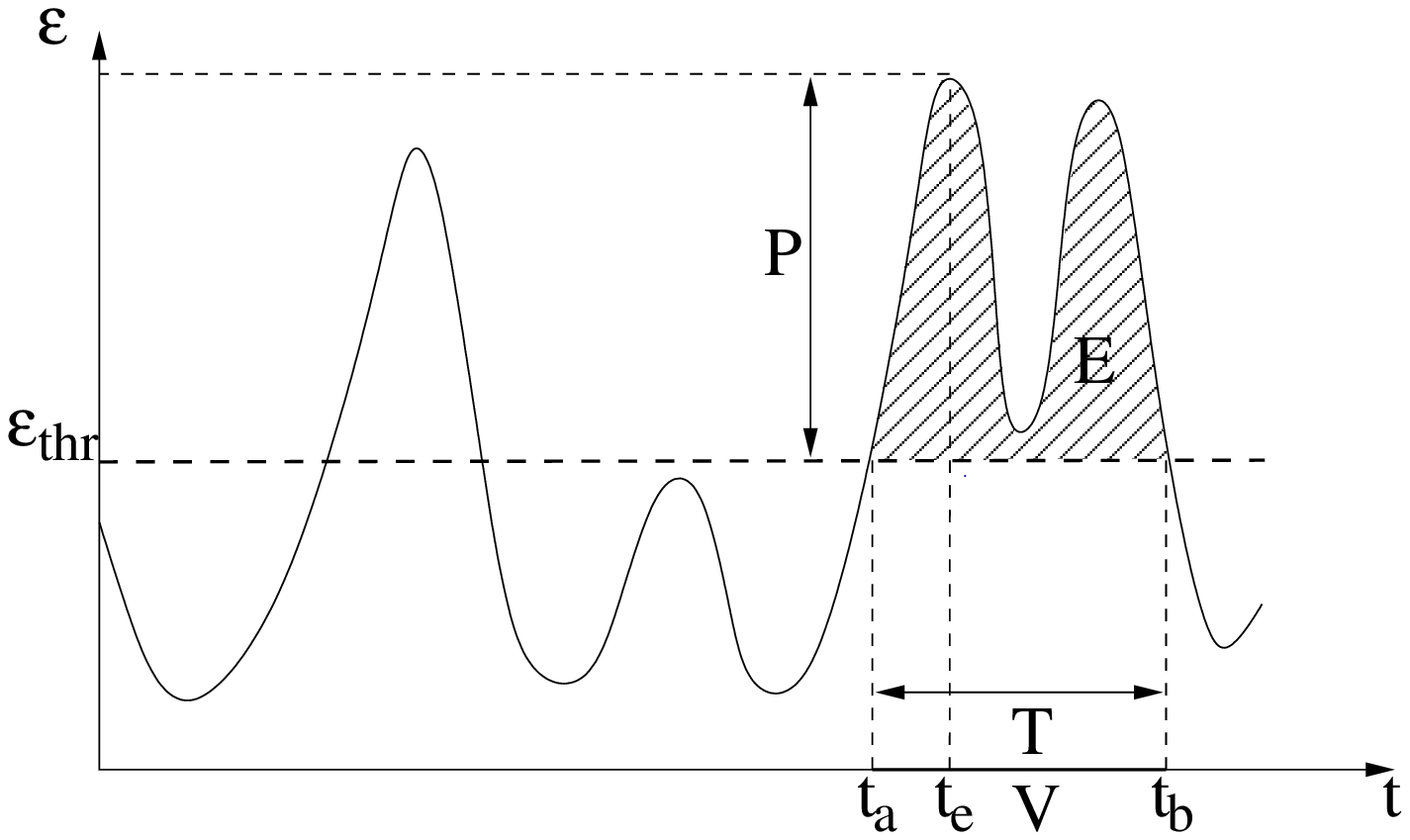}\\
    \includegraphics[width=\linewidth]{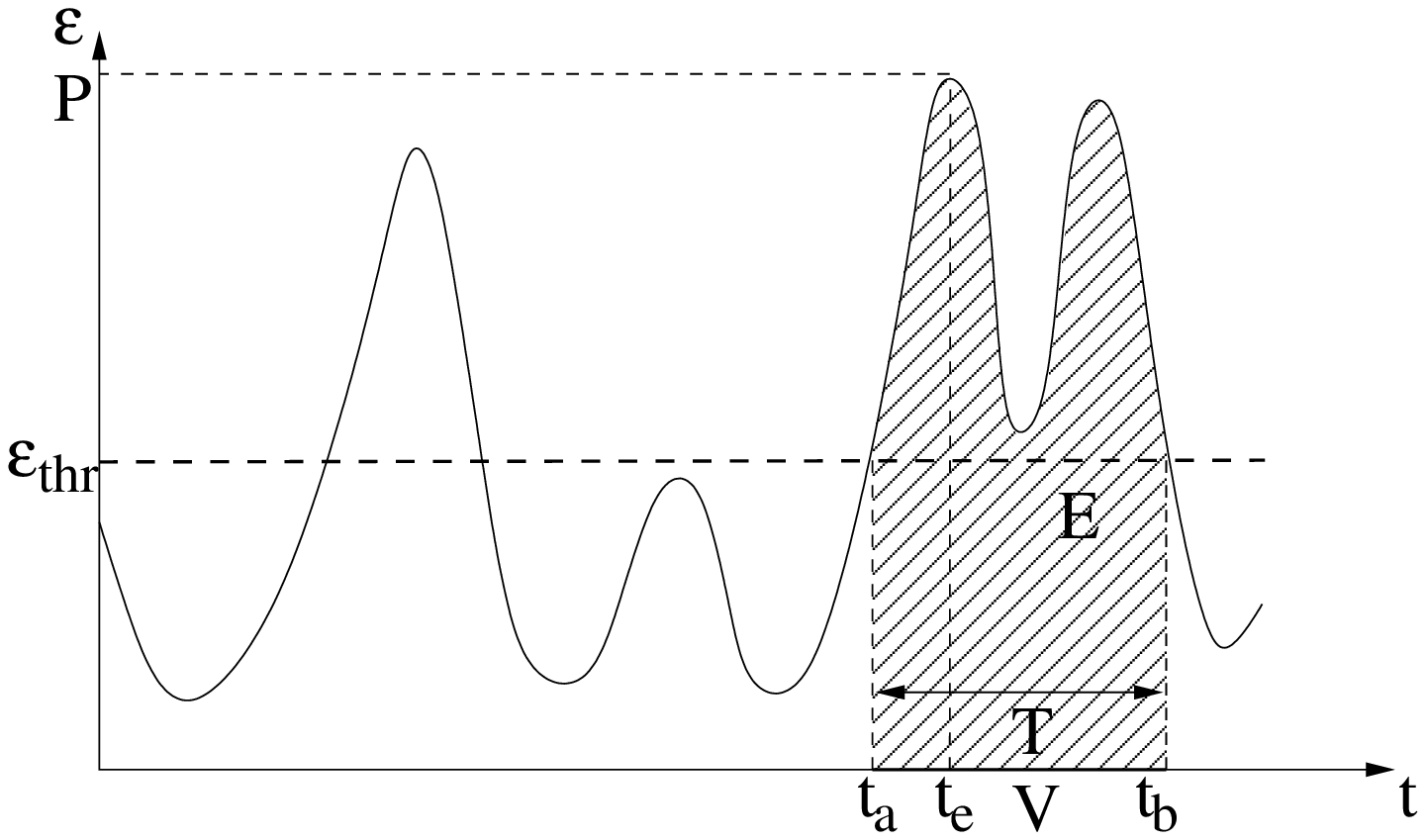}
  \end{minipage}
  \begin{minipage}{.33\linewidth}
    \includegraphics[width=\linewidth]{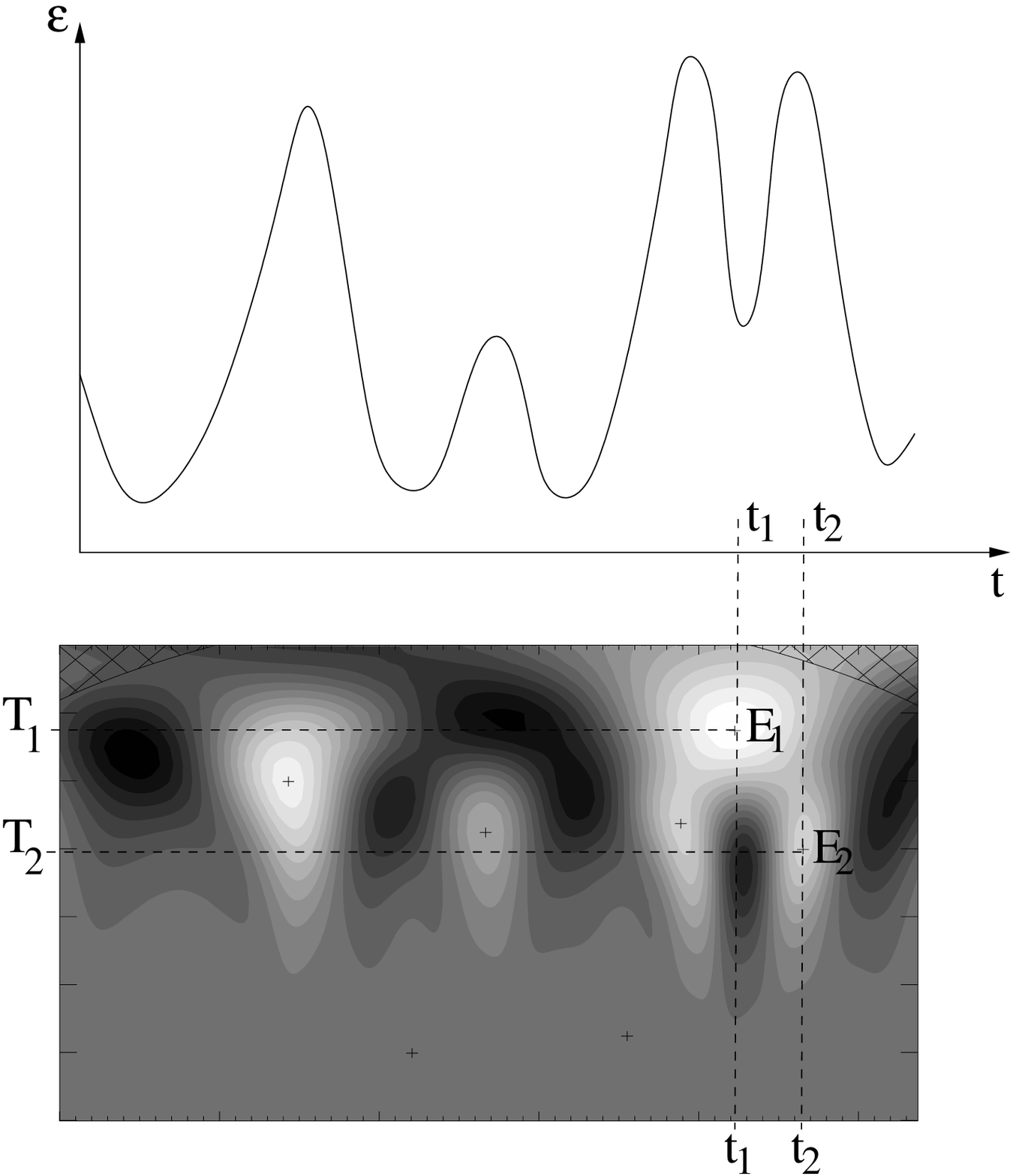}
  \end{minipage}
  \caption{Definitions of events and of event characteristics ($t_e$, $T$,
    $P$, and $E$). Left: peaks (definition \ref{def:peak}), with (top) and
    without (bottom) background detection.  Middle: threshold (definition
    \ref{def:thr}), with (top) or without (bottom) taking the background
    into account.  Right: peaks in the wavelets time-frequency plane
    (definition \ref{def:wave}). For this definition, events are marked as
    crosses in the time-frequency plane, indicating the event time $t_e$ and
    duration $T$. The total energy $E$ is the wavelet amplitude (color) at
    this position.}
  \label{fig:defs}
\end{figure*}

We present here a basic list of possible definitions of an event when a
``signal'' $\epsilon(t)$, which is the time series of the power dissipated
in the system, is given (Fig.~\ref{fig:defs}). Most of the ideas of this
list come from the definitions that have been chosen in papers
found in the literature. For each event, we get:
\begin{itemize}
\item $E$, the total energy dissipated during the event,
\item $P$, the peak power of energy dissipation,
\item $T$, the duration of the event,
\item $t_e$, the time of the event, necessary to obtain the waiting times
  $\tau_w$, \ie the (quiescent) time between two consecutive events.
\end{itemize}

\subsection{Peaks}

\begin{defin}[peak]
  \label{def:peak}
  An event corresponds to a local maximum $\epsilon(t_m)$ in the signal
  $\epsilon(t)$. The time of the event is $t_e = t_m$, the peak dissipation
  power is $P = \epsilon(t_e)$, the total dissipated energy is $E =
  \int_{t_a}^{t_b} \epsilon(t) \ud t$ where $t_a$ and $t_b$ are the two
  local minima around $t_e$, and the event duration is $T = t_b - t_a$.
\end{defin}

\begin{variant}[peak-background]
  \label{def:peakbg}
  The background $b(t)$ is the affine function defined between the points
  $(t_a, \epsilon(t_a))$ and $(t_b, \epsilon(t_b))$. With the notations of
  definition \ref{def:peak}, the time of the event is $t_e$, the peak
  dissipation power is $P-b(t_e)$, the total dissipated energy is $E -
  \int_{t_a}^{t_b} b(t) \ud t = E - (\epsilon(t_a) + \epsilon(t_b)) \cdot T
  / 2$, and the event duration is $T$.
\end{variant}

\subsection{Threshold}

\begin{defin}[threshold]
  \label{def:thr}
  A threshold $\epsthr$ is chosen, and an event is a part of the signal
  $\epsilon(t)$ which stays above $\epsthr$: more precisely, it is a maximal
  connex part $V=[t_a, t_b]$ of the set $\{t \;|\; \epsilon(t) > \epsthr\}$.
  The total dissipated energy is $E = \int_V \epsilon(t) \ud t$, the peak
  dissipation power is $P = \max_V \epsilon(t)$ and the event duration is $T
  = t_b - t_a$. The time of the event is the time at which the maximum of
  $\epsilon(t)$ on $V$ is attained: $\epsilon(t_e) = \max_V \epsilon(t)$.
\end{defin}

There are several other alternatives to define the time of the event, like
$t_e = (t_b + t_a) / 2$ (the middle of interval $V$) or $t_e = \frac{1}{E}
\int_V \epsilon(t) \cdot t \ud t$ (the barycenter of the event, weighted by
$\epsilon$). But these variants do not change the statistics of $P$, $E$,
and $T$, and they seem to have little influence on the statistics of
$\tau_w$.

We can consider the threshold level $\epsthr$ as a background level, giving
the following definition:

\begin{variant}[threshold-background]
  \label{def:thrbg}
  Using definition \ref{def:thr} and its notations, the time of the event is
  $t_e$, the peak dissipation power is $P-\epsthr$, the total dissipated
  energy is $E - \int_V \epsthr \ud t = E - \epsthr \cdot T$, and the event
  duration is $T$.
\end{variant}

The threshold can be chosen as a function of the overall average
$\bar\epsilon$ and standard deviation $\sigma_e$ of $\epsilon(t)$. It can
also be chosen iteratively, by using the average and standard deviation of
the time series during the quiescent times between events (which have been
defined by the previous iteration of this process), as in \citet{bof99}.

\subsection{Wavelet analysis}

This method assumes that we have built the time-scale plane $y(t_0,s)$ for
$\epsilon(t)$, by convolution of $\epsilon(t)$ by the wavelets $w_{t_0, s}
(t) = 1/s \cdot w_0((t-t_0)/s)$. A mother wavelet $w_0$ which seems well
adapted to the shape of events is the second derivative of a Gaussian
(``Mexican hat''). When the noise is in $1/f$, \citet{san01} have shown that
the Mexican hat is the best wavelet to find enhancements of the signal.

\begin{defin}[wavelet]
\label{def:wave}
An event corresponds to a local maximum $y(t_e, s_e)$ in the time-scale
plane $y(t_0, s)$. The time of the event is $t_e$, its duration $T$ is the
scale $s_e$, its total energy is $E = y(t_e, s_e)$. Its peak power $P$ can
be defined as $\max_V \epsilon(t)$ with $V = [t_e-s_e/2,\,t_e+s_e/2]$.
\end{defin}

For better accuracy in the definition of $s_e$ and $t_e$, we need to have a
good resolution in the time-frequency plane, \ie we use a continuous wavelet
transform. As a result, and also because we have used the Mexican hat
wavelet, the wavelets are non-orthogonal, and there is a risk to find
several non-independent events where there is only one event. To avoid this,
it is possible to impose that two events are separated by at least some
distance in the time-scale plane (typically a factor $2$ in scale and an
interval $s_e$ in time), but in practice this is not necessary as the
time-scale plane is sufficiently smooth.

\section{Model time series}
\label{sec:modelts}

\subsection{Shell-model}

\label{sec:shell}
The results of the next section are based on data output from a shell-model
of incompressible MHD turbulence. In such models \citep{glo85, bis94, fri98,
  giu98}, the Fourier wavenumber space is divided in concentric shells $S_n
= \{\vec{k} \;|\; \|\vec{k}\| \in [k_n, k_{n+1}]\}$ with $k_n = k_0 \lambda
^ n $ and usually $\lambda=2$. A single complex scalar value $u_n$
represents the velocity increments $|u(x+\ell)-u(x)|$ on scales $\ell$ for
$2\pi/\ell \in S_n$. The same holds for the scalar value $b_n$ representing
the magnetic field increments on the same scales $\ell$.  This model is the
magnetohydrodynamic analog of the GOY \citep[Gledzer-Ohkitani-Yamada:
][]{gle73,yam87,yam88,yam88b} shell-model of fluid turbulence. It is
governed by the following equations, given in \citet{giu98}:
\begin{equation}
\label{eq:dtz}
\frac{\mathrm{d}Z_n^s}{\mathrm{d}t} = -k_n^2 (\nu^+ Z_n^s + \nu^-
Z_n^{-s}) + i k_n T_n^{s*} + f_n^s
\end{equation}
where $Z_n^s=v_n+s b_n$ are the Els\"asser variables, $s = \pm$,
$\nu^s = (\nu+s\eta)/2$ are combinations of kinematic viscosity and
resistivity, $f_n^s$ are external driving forces, and
$T_n^s$
is the term corresponding to local non-linear interactions between shells.
For a given shell $n$, this term involves the neighbors
and second-nearest neighbors of the shell $n$, modelling local interactions
between triads of consecutive modes. The detailed coefficients of this term
are given in \citet{giu98} and depend on the dimensionality (\eg 2D or 3D)
of the physical MHD system the shell-model represents, \emph{via} the
conservation of the MHD invariants.


This model can describe the evolution of modes over a wide range of
wavenumbers with just a few dozens of degrees of freedom. It is thus very
interesting for studying MHD turbulence with high Reynolds numbers, and
intermittency. It actually exhibits typical properties of MHD turbulence,
from wide power-law spectra to dynamo effect in 3D \citep{giu98}, including
spatial and temporal intermittency \citep{giu98,bof02}.

The equations \ref{eq:dtz} are solved numerically and we get the time series
of dissipated power $\epsilon(t) = \sum_n k_n^2 (\nu |u_n(t)|^2 + \eta
|b_n(t)|^2)$, which is our variable of interest. To obtain the first time
series shown on Fig.~\ref{fig:ts}, hereafter known as the time series
\tise{1}, we used 24 shells (representing $k=1$ to $k\approx 8.4\,10^6$),
with $\lambda=2$, $\nu=\eta=10^{-11}$. We performed $10^{7}$ variable
timesteps (determined by a CFL condition) with a 3$^\textind{rd}$-order
Runge-Kutta numerical scheme. The time series \tise{2}, also shown on
Fig.~\ref{fig:ts}, was obtained with the same parameters, except that the
dissipation coefficients $\nu$ and $\eta$ where ten times higher than for
\tise{1}.

\subsection{Coupled shell-models}

\label{sec:sl}
Section \ref{sec:int} uses also data from a version of a shell-model
designed to model a region of space where a dominant magnetic field
$\vec{B}_0$ exists, like in a coronal loop \citep{buc04soho15}. In this
model, shell-models of 2D MHD are coupled by Alfvén waves travelling along
$\vec{B}_0$, and energy is only input by movements of the photospheric
footpoints of the loop. This geometric setup is the same as the one used for
the cellular automaton described by \citet{buc03}, and it gives a model
similar to the one described by \citet{nig04}. Here we use an independent
implementation of these ideas to obtain the time series \tise{3}.

\subsection{Characteristics and intermittency of the time series}

\begin{figure}[tbp]
  \includegraphics[width=\linewidth]{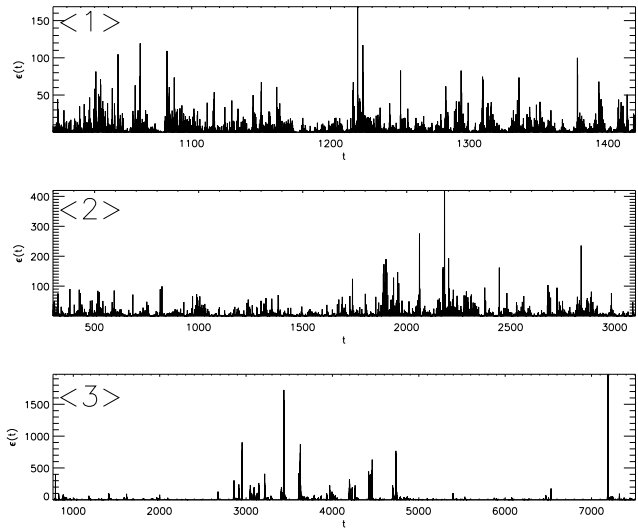}
  \caption{From top to bottom, time series \tise{1}, \tise{2},
    and \tise{3}.}
  \label{fig:ts}
\end{figure}

\begin{figure}[tbp]
\includegraphics[width=\linewidth]{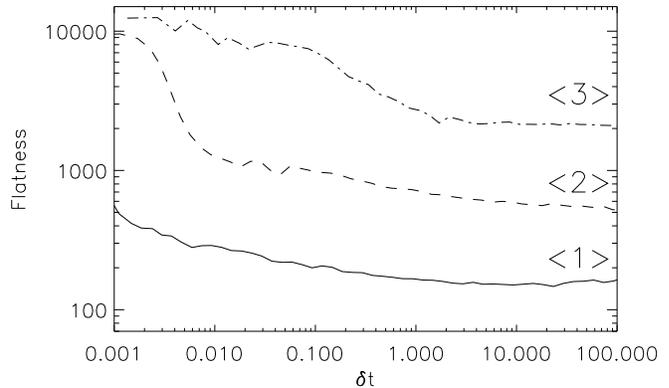}
\caption{The flatness of the three time series increases when
  the scale $\delta t$ decreases. This behavior is a signature of
  intermittency, and is stronger for the time series \tise{3} and lower for
  time series \tise{1}. Note that the flatness remains much higher than the
  Gaussian value $3$ even at large time scales, as a consequence of the
  non-Gaussian distribution function of the values taken by the time
  series.}
\label{fig:flat}
\end{figure}

All these time series were rescaled so that their average $\bar{\epsilon}$
is $1$ and are shown on Fig\ \ref{fig:ts}. Their basic characteristics are
summarized in table \ref{tab:charts}. From time series \tise{1} to \tise{3},
the ratio of the maximum (or the standard deviation) to the average grows,
and longer, quiet times exist between the intervals with higher dissipation.
It seems that intermittency is higher for \tise{3} than for \tise{2}, and
that it is also higher for \tise{2} than for \tise{1}. This is verified by
plotting the flatness\footnote{We use the following definition for the
  flatness $F(\tau)$ of the time series $\epsilon(t)$:
  $F(\tau)=S^4(\tau)/(S^2(\tau))^2$, where $S^q(\tau)=\langle
  |\epsilon(t+\tau) - \epsilon(t)|^q\rangle_t$ is the structure function of
  index $q$ for the timeseries $\epsilon$.} of these time series as a
function of the temporal scale (Fig.~\ref{fig:flat}).

\begin{table}[tbp]
  \centering
  \caption{Summary of the characteristics of all 3 time series, which were
    normalized so that their average is $\bar\epsilon=1$: number of
    data points, number of peaks, standard deviation, and maximum value.}
  \label{tab:charts}
  \begin{tabular}{ccccc}
    \hline\hline
    & Data points & Peaks & $\sigma_\epsilon$ & $\epsilon_\textind{max}$ \\
    \hline
    \tise{1} & 453,628 & 51,507 & 1.98 & 169  \\
    \tise{2} & 985,162 & 56,136 & 2.45 & 420 \\
    \tise{3} & 1,000,000 & 305,738 & 11.33 & 1971  \\ \hline
  \end{tabular}
\end{table}

\section{Comparison between statistics for different event definitions}

In this section, starting from the time series $\epsilon(t)$ number \tise{1}
produced as explained in section \ref{sec:shell}, we compare the effect of
the definition of an event for the following statistics:
\begin{itemize}
\item normalized histograms (\ie experimental Probability Distribution
  Functions -- or PDFs) of event peak dissipation power, total energy and
  duration, as defined by the different event definitions we use,
\item PDFs of waiting times between events, \ie the time between two
successive events. This corresponds to the laminar, quiet time between
events.
\end{itemize}

\subsection{Peaks}

As each peak of the time series is counted as an event, definitions
\ref{def:peak} and \ref{def:peakbg} give a lot of events, even in the case
of our numerical data, which has no noise: for time series \tise{1} for
example, one data point over nine is a local maximum, and corresponds thus
to an event. When noise is present, a smoothening of the data at the scale
of the shortest events may be necessary before searching for events.
Furthermore, the set of the events is a partition of the time series (the
end $t_b$ of one event is the beginning $t_a$ of the next event), all the
energy of the time series is contained in the events: $\sum_i E_i=\int
\epsilon(t)\,\ud t$.

\begin{figure}[tbp]
  \includegraphics[width=\linewidth]{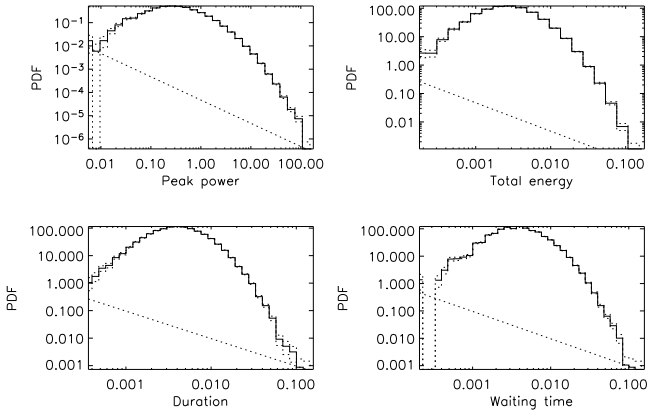}
  \caption{Events statistics of time series \tise{1} for definition
    \ref{def:peak} (peaks): peak power of energy dissipation, dissipated
    energy, duration, and waiting times. The straight dashed line
    corresponds to one event per histogram bar; as the histogram bars are
    spaced exponentially, its slope is -1. The dashed histograms are an
    estimation of the discretization error when building the histogram,
    computed assuming Poisson statistics in each bar ($\pm\sqrt{N}$
      where $N$ is the number of events in a given histogram bar).}
  \label{fig:statpeak}
\end{figure}

\begin{figure}[tbp]
  \centering
  \includegraphics[width=.6\linewidth]{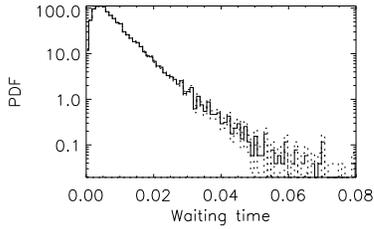}
  \caption{Waiting-time distribution of events from time series \tise{1} for
    definition \ref{def:peak} (peaks), in lin-log scale.}
  \label{fig:statpeakwt}
\end{figure}

The distributions of $P$, $E$, $T$ and $\tau_w$ (Fig.~\ref{fig:statpeak})
have approximately the same shape, which is neither a power-law, nor an
exponential or gaussian. The tail of the waiting-time distribution could
even be fitted to an exponential (Fig.~\ref{fig:statpeakwt}), in
contradiction with previous studies of shell-models, which used another
definition of an event \citep{bof99,lep01}.

With these definitions, even the smallest peaks are counted as events, and
this breaks the waiting times into small parts, leading to a cut-off of the
tail of the waiting-time distribution. To decrease this effect, we may
exclude the smallest events (\eg those with a peak power lower than a given
threshold) from the analysis, which gives the following variant of
definition \ref{def:peak}:

\vspace{1ex} \textbf{Variant 1.2 (peak-threshold).}  A threshold $\epsthr$
is chosen. An event corresponds to a local maximum $\epsilon(t_m)$ in the
signal $\epsilon(t)$, \emph{provided that $\epsilon(t_m) > \epsthr$}. The
time of the event is $t_e = t_m$, the peak dissipation power is $P =
\epsilon(t_e)$, the total dissipated energy is $E = \int_{t_a}^{t_b}
\epsilon(t) \ud t$ where $t_a$ and $t_b$ are the two local minima around
$t_e$, and the event duration is $T = t_b - t_a$.  \vspace{1ex}

Note that this is \emph{not} the same as using definition \ref{def:thr}: for
a given threshold $\epsthr$, on a maximum connex part $V$ of $\{t \;|\;
\epsilon(t) > \epsthr\}$, definition 1.2 will find as many events as there
are peaks of $\epsilon(t)$ on the interval $V$, whereas definition
\ref{def:thr} will find only one event.

\begin{figure}[tbp]
  \centering
  \includegraphics[width=\linewidth]{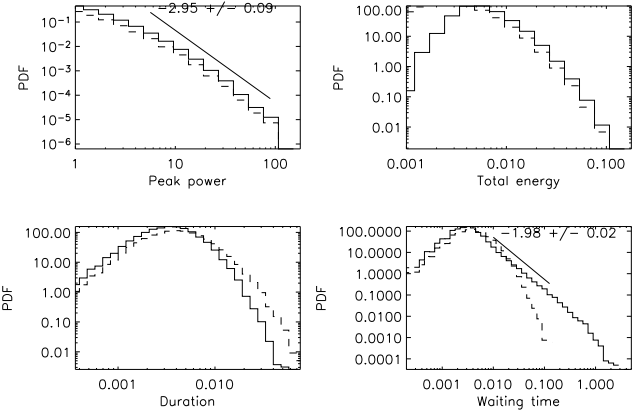}
  \caption{
    Events statistics of time series \tise{1} for definition 1.2
    (peak-threshold) with $\epsthr=1$ (average of time series). The
    statistics for definition \ref{def:peak} (peaks) are shown as a
    reference, in dashed lines.  }
  \label{fig:statpeakthr}
\end{figure}

As a result (Fig.~\ref{fig:statpeakthr}), it is clear that the PDF of $P$ is
cut below the value of $\epsthr$, with no modification of its shape: this
means that only the tail $P > \epsthr = 1$ of the histogram of $P$ in
Fig.~\ref{fig:statpeak} is left, and this tail could be fitted to a quite
narrow power-law of slope $-2.95$.  The PDFs of $E$ and $D$ do not change
dramatically, except that the left part is weaker because of the
correlations between $P$, $E$ and $D$.  The most interesting effect of using
variant 1.2 instead of definition \ref{def:peak} is for the waiting-time
distribution: it exhibits now a clear power-law of index $-1.98$ over $2.5$
decades. This is made possible by the fact that small waiting times
associated to small events in the case of definition \ref{def:peak} are now
replaced by a smaller number of long waiting times, leading to a
reinforcement of the right part of the histogram of $\tau_w$.

\subsection{Threshold}

\begin{figure}[tbp]
  \includegraphics[width=\linewidth]{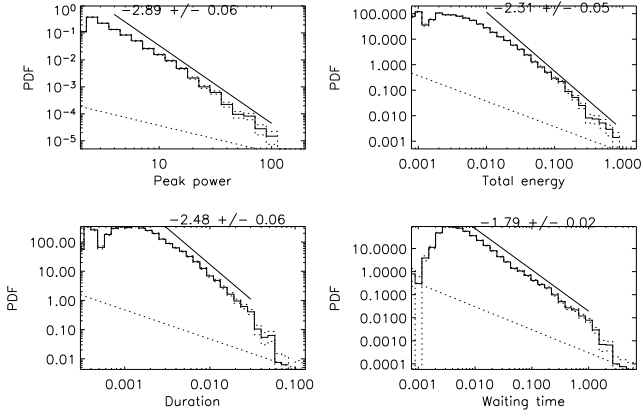}
  \caption{Events statistics of time series \tise{1} for definition
    \ref{def:thr} (threshold).}
  \label{fig:statthr}
\end{figure}

With definition \ref{def:thr}, histograms of $P$, $E$, $T$, and $\tau_w$ are
quite clearly power-laws (Fig.~\ref{fig:statthr}), even if they are not very
wide for this weakly intermittent time series. The slopes of these
power-laws are $2.89\pm0.06$ for $P$, $2.31\pm0.05$ for $E$, $2.48\pm0.06$
for $T$, and $1.79\pm0.02$ for $\tau_w$.

\begin{figure}[tbp]
  \includegraphics[width=\linewidth]{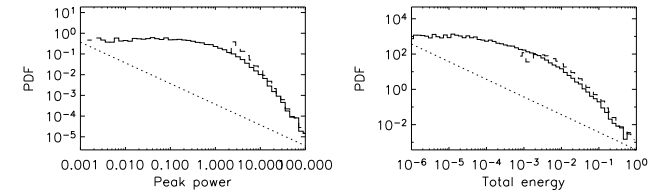}
  \caption{Events statistics of time series \tise{1} for definition
    \ref{def:thrbg} (threshold-background). Only the distributions which are
    different from the one obtained by definition \ref{def:thr} (threshold,
    Fig.~\ref{fig:statthr}) are represented.}
  \label{fig:statthrbg}
\end{figure}

These power-law tails still exist when the threshold is considered as a
background and is removed (variant \ref{def:thrbg},
Fig.~\ref{fig:statthrbg}), but the left part of the histograms is then
almost flat in logarithmic axes.  This is for example quite straight-forward
for the distribution of $P$, as removing the background is shifting --- in
linear axes --- the distribution of $P$ to the left. However, the right tail
of the distributions, \eg for $P \gg \epsthr$ remains almost the same when
the background is removed.  It seems that background removal does not help
understand the statistics of events.

Methods using a threshold are very widely used when events are searched in
time series, as well from numerical simulations: \citet{dmi98,ein96a,gev98}
(2D RMHD), \citet{bof99} (MHD shell-models) as from X-rays observations:
 \citet{pea93,cro93,whe98}. It actually
seems to be well-adapted to instrumental constraints of sensitivity and
noise levels. 

The drawbacks of this definition are that it misses the lowest-energy events
(leading to a cut-off of the left part of the energy histograms), and that
it cannot separate close high-energy events.  This definition is also not
  adapted to non-stationary time series: in this case, the threshold should
  adapt to the local statistical characteristics of the time series.

\subsection{Wavelet analysis}

\begin{figure}[tbp]
  \includegraphics[width=\linewidth]{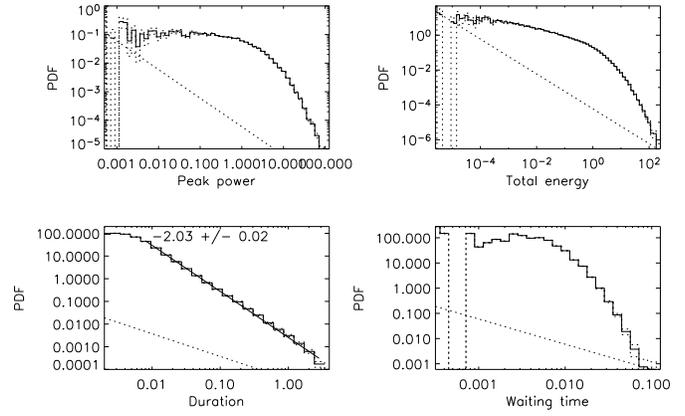}
  \caption{Events statistics of time series \tise{1} for
    definition \ref{def:wave} (maxima in wavelet time-scale space).}
  \label{fig:stattf}
\end{figure}

Definition \ref{def:wave} produces the histograms of $P$, $E$, $D$, and
$\tau_w$ shown on Fig.~\ref{fig:stattf}. The histogram of events durations
is a power-law over more than $2.5$ decades. The wide and flat left part of
the histograms of $P$ and $E$, which include events much smaller than with
other definitions, suggest to use a variant of definition \ref{def:wave}
similar to variant 1.2 of definition \ref{def:peak}, where the smallest
events are simply not taken into account:

\begin{variant}[wavelet-threshold]
  \label{def:wavethr}
  A threshold $E_\textind{thr}$ is chosen.  An event corresponds to a local
  maximum $y(t_e, s_e)$ in the time-scale plane $y(t_0, s)$, \emph{provided
    that $y(t_e, s_e) > E_\textind{thr}$}. The time of the event is $t_e$,
  its duration $T$ is the scale $s_e$, its total energy is $E = y(t_e,
  s_e)$.  Its peak power $P$ can be defined as $\max_V \epsilon(t)$ with $V
  = [t_e-s_e/2,\,t_e+s_e/2]$.
\end{variant}

\begin{figure}[tbp]
  \includegraphics[width=\linewidth]{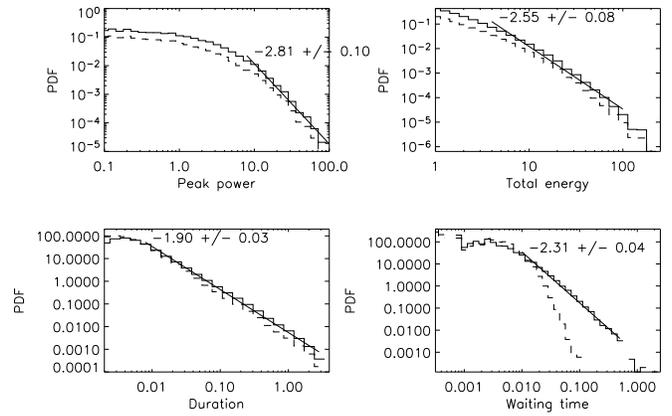}
  \caption{Events statistics of time series \tise{1} for
    definition \ref{def:wavethr} (wavelet-threshold) and $E_\textind{thr}=1$.
    The statistics for definition \ref{def:wave} (maxima in wavelet time-scale
    space) are shown as a reference, in dashed lines.}
  \label{fig:stattfthr}
\end{figure}

As for definition 1.2 (peak-threshold), the distributions of $P$, $E$ and
$D$ do not change much, but a power-law is recovered
(Fig.~\ref{fig:stattfthr}) for the waiting-time distribution.

\section{Intermittency and sensitivity to event definition}

\label{sec:int}

\subsection{Sensitivity to event definition}

\begin{figure}[tbp]
  \centering
  (a)~\tise{1}\hspace{.42\linewidth}(b)~\tise{2}\\[-2mm]
  \includegraphics[width=.49\linewidth]{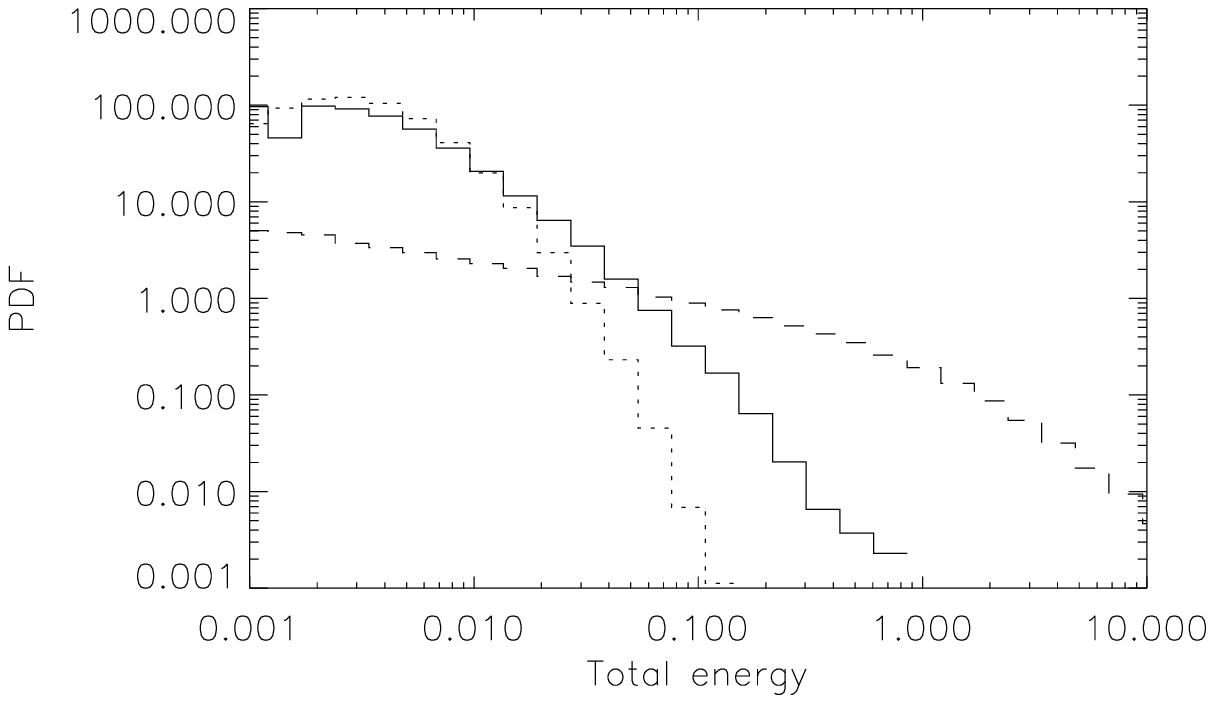}
  \includegraphics[width=.49\linewidth]{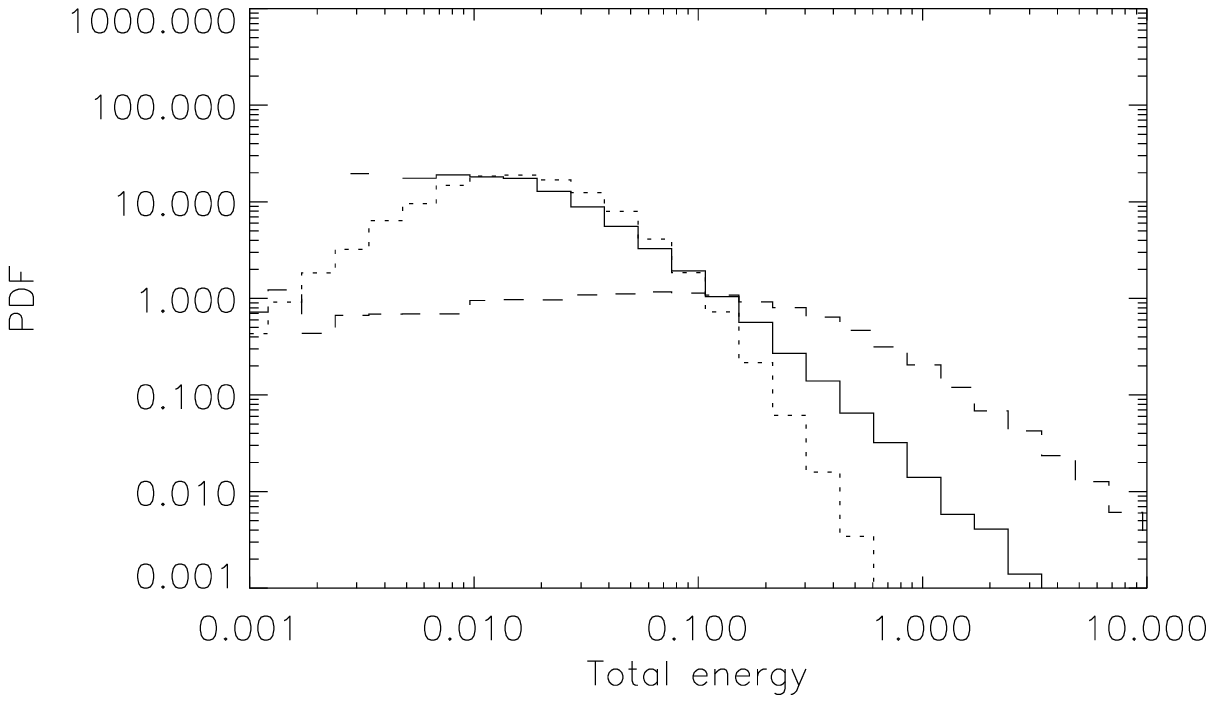}\\
  (c)~\tise{3}\\[-2mm]
  \includegraphics[width=.49\linewidth]{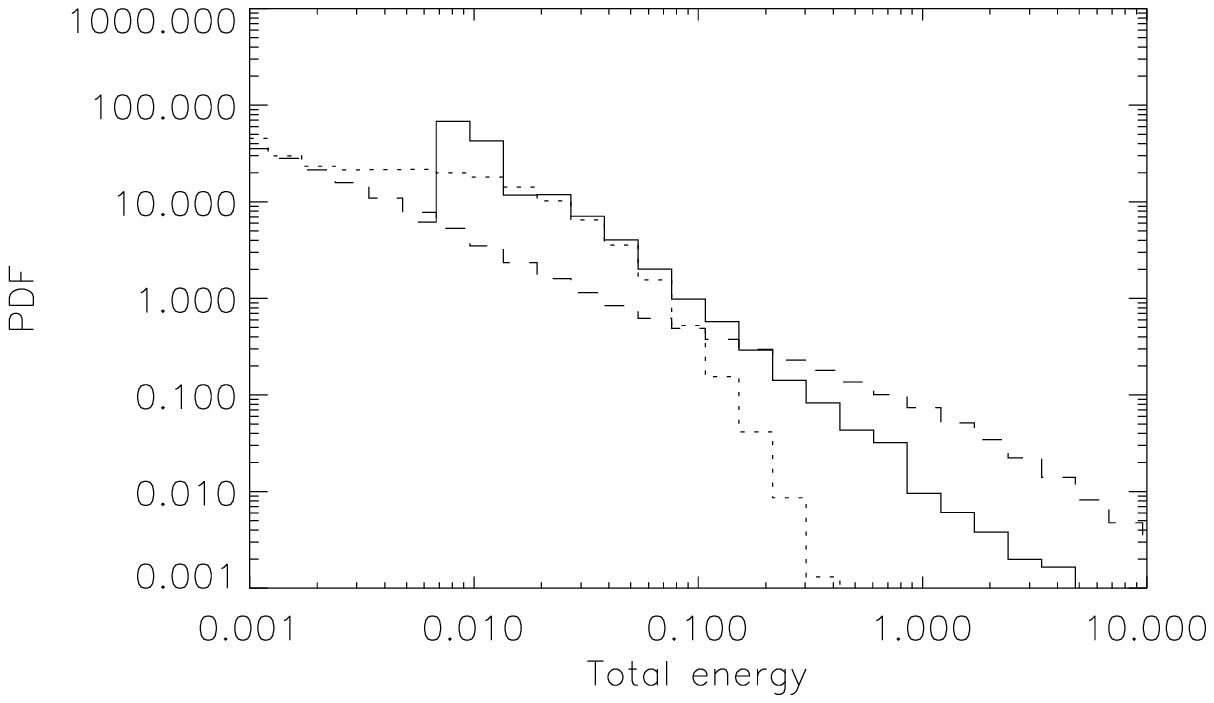}
  \caption{(a) Total energy distributions for events
    determined by definitions \ref{def:peak} (peaks; dotted line),
    \ref{def:thr} (threshold; plain line), and \ref{def:wave} (maxima in
    wavelet time-scale space; dashed line), for time series \tise{1}. (b)
    Same figure for time series \tise{2}. (c) Same figure for time series
    \tise{3}. All plots have the same scale. }
  \label{fig:inten}
\end{figure}

\begin{figure}[tbp]
  \centering
  (a)~\tise{1}\hspace{.42\linewidth}(b)~\tise{2}\\[-2mm]
  \includegraphics[width=.49\linewidth]{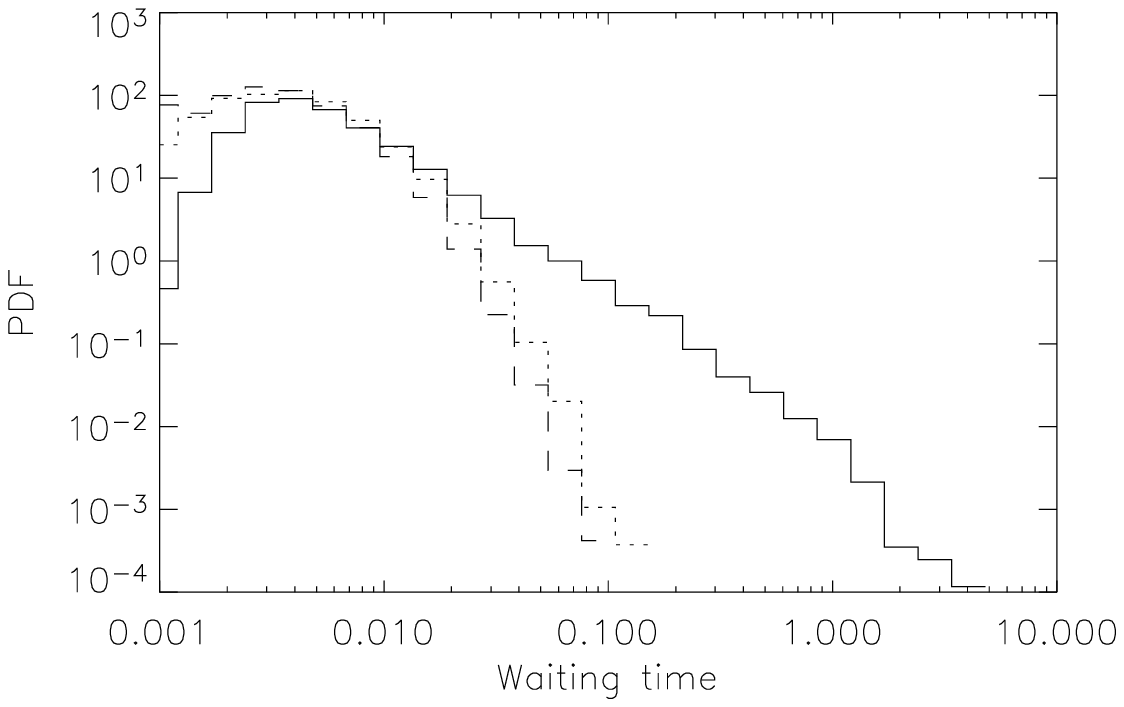}
  \includegraphics[width=.49\linewidth]{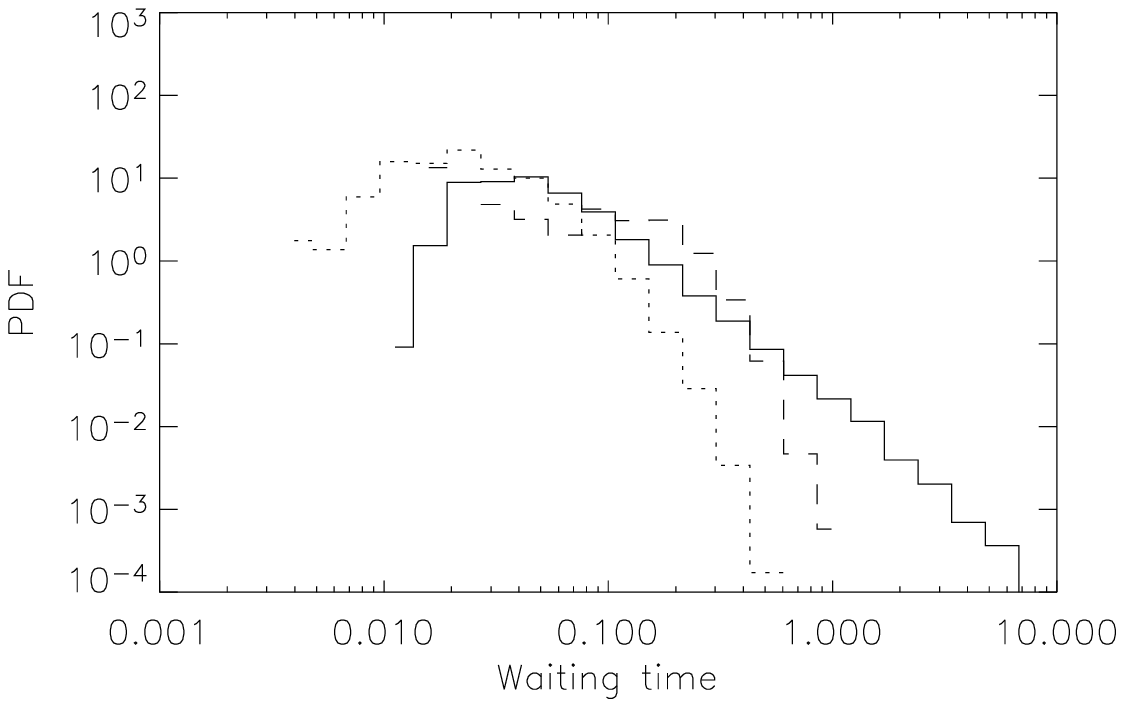}\\
  (c)~\tise{3}\\[-2mm]
  \includegraphics[width=.49\linewidth]{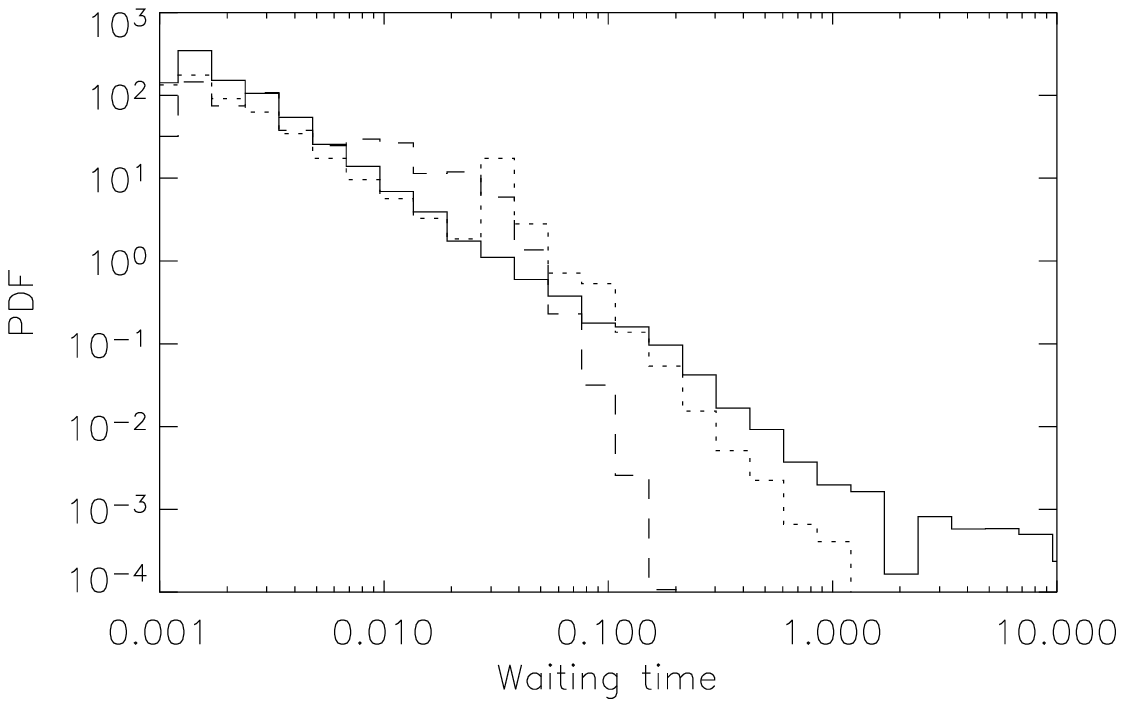}
  \caption{(a) Waiting-time distributions for events
    determined by definitions \ref{def:peak} (peaks; dotted line),
    \ref{def:thr} (threshold; plain line), and \ref{def:wave} (maxima in
    wavelet time-scale space; dashed line), for time series \tise{1}. (b)
    Same figure for time series \tise{2}. (c) Same figure for time series
    \tise{3}. All plots have the same scale. }
  \label{fig:wt}
\end{figure}

Now we use all the 3 time series described in section \ref{sec:modelts}. It
seems that the distributions of event energies $E$ obtained by definitions
\ref{def:peak} and \ref{def:wave} are closer from the power-law obtained by
definition \ref{def:thr} in the case of time series \tise{3}
(Fig.~\ref{fig:inten}c) than in the case of time series \tise{1}
(Fig.~\ref{fig:inten}a). The waiting-times distributions (Fig.~\ref{fig:wt})
display the same behavior. In general, distributions obtained from
higher-intermittency time series seem to be less sensitive to the definition
of an event than low-intermittency time series.

\subsection{Sensitivity to threshold, for definition \ref{def:thr}
  (threshold)}

\begin{figure}[tbp]
  \centering
  \includegraphics[width=\linewidth]{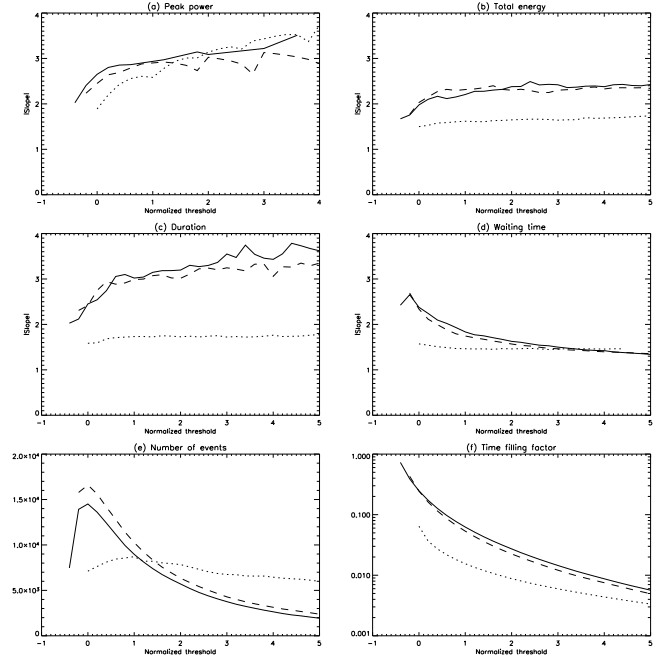}
  \caption{
    Slope of the peak power (a), total energy (b), duration (c), and waiting
    time (d) distributions, for events defined by definition \ref{def:thr}
    (threshold), as a function of the normalized threshold $(\epsthr -
    \bar{\epsilon}) / \sigma_\epsilon$. The plain, dashed, and dotted lines
    correspond to time series \tise{1}, \tise{2}, and \tise{3} respectively.
    (e) Number of events detected as a function of the normalized threshold.
    (f) Proportion of the duration contained in events, as a function of the
    normalized threshold.}
  \label{fig:thrslope}
\end{figure}

In the case of events defined by a threshold
  (like definition \ref{def:thr}), the slope of event energy histograms may
  depend on the threshold. Here we choose different values of $\epsthr$
between $0$ and $\bar\epsilon + 5\sigma_\epsilon$ where $\bar\epsilon$ is
the time series average ($\bar\epsilon=1$) and $\sigma_\epsilon$ is the
standard deviation shown in table \ref{tab:charts} for each of the time
series. As a result, the number of events (Fig.~\ref{fig:thrslope}e) is $1$
when the threshold is $\epsthr=0$ (the whole time series is \emph{one}
event); it increases to a maximum, attained between $\bar\epsilon$ and
$\bar\epsilon + \sigma_\epsilon$, depending on the time series
characteristics; then it decreases (ultimately, the number of events is $0$
when $\epsthr>\epsilon_\textind{max}$, where $\epsilon_\textind{max}$ is the
maximum value of the time series).

Figure\ \ref{fig:thrslope} shows the power-law slope of the histograms of
$P$, $E$, $T$, and $\tau_w$ as a function of the normalized threshold
$(\epsthr - \bar{\epsilon}) / \sigma_\epsilon$. In general, time series
\tise{1} and \tise{2}, which come from the same simple shell-model and which
are less intermittent than the time series \tise{3}, follow quite the same
path. (a) The distributions of peak dissipation power $P$ have a slope
$\approx 2$ for a low threshold $\epsthr$, and become steeper when $\epsthr$
increases. The slope for time series \tise{3} is slightly more sensitive to
$\epsthr$ than the other time series. (b) The slope of the distributions of
energy $E$ also increase with $\epsthr$, except for time series \tise{3},
for which it is almost constant. (c) The statistics of the durations $T$
exhibit the same features than the statistics of $E$. (d) On the contrary,
the slope of the distributions of the waiting times $\tau_w$ decreases when
$\epsthr$ increases. Again, it is almost constant for time series
\tise{3}. (f) The proportion of the time series duration contained
  in events decreases when the threshold increases. This decrease is
  stronger for the lowly intermittent time series.

Time series \tise{3} seems to be the least sensitive to the value of
$\epsthr$. Note that, by using thresholds expressed as a function of
$\sigma_\epsilon$ instead of absolute thresholds, we have taken care of the
fact that the deviations of time series \tise{3} are larger than for the
other time series.

\section{Discussion}

We have investigated the dependence on the definition of ``events'' of the
statistics of events obtained from an energy dissipation time series. Not
very surprisingly, the statistics of peak power, energy content, duration
and waiting times of events differ when different definitions are used.

Especially for low-intermittency time series and for waiting-time
distributions, power-law distributions are recovered only when a threshold
is used, either when searching for events (definition \ref{def:thr}) or
after having searched for events by another means (definitions 1.2 and
\ref{def:wavethr}). It is also interesting to note that the waiting-time
distribution, which is used to test the Poissonian nature of the flaring
process \citep{whe98,lep01,whe02}, can have a power-law or an exponential
tail, depending on the definition of events.

For observational studies, where the smallest events are averaged over the
line of sight and the spatial and temporal steps, some of the intermittency
is lost. In this case we need to use a definition which gives statistics as
close as possible from the statistics of the underlying (non-averaged)
signal (which is intermittent enough for events statistics to be almost
independent from their definition). The definitions using a threshold seem
to be adequate from this point of view.  The presence of noise in
observations gives also a strong support to this kind of definitions.
However, these definitions have also drawbacks, in particular the difficulty
of choosing a threshold for a non-stationary time series.

Other definitions like \ref{def:wave} which uses wavelets can have
interesting properties separating simultaneous events at different scales,
but the smallest events obtained by this means seem to be not significant.
Alternatively, events could be defined iteratively from the time-scale
plane: the first event is defined by the overall maximum of the time-scale
plane, the corresponding wavelet is subtracted from $\epsilon(t)$, a new
time-scale plane is computed, and this process is done again to find each of
the next events. Local Intermittency Measure \citep[LIM:][]{farge90} could
perhaps also be used for this purpose. However, these ideas have not been
investigated further yet and an iterative definition may be computationaly
very expensive compared to the other definitions.

Let us now return to the motivation behind the determination and discussion
of event properties and statistics for coronal physics.  There are two main
reasons for these studies, essentially related to bridging the gap between
observable time and spatial scales and the sub-resolution physics.  On the
one hand, we would like to understand whether analogous physical processes,
namely flares, conserve scale-invariant properties at unobservable scales
and are responsible for the existence of the quiet corona as we know it. On
the other, one would like to link, as far as possible, large scale physical
models and numerical simulations to the observations without reproducing in
detail the microcosm of a single small-scale event (though this may be
desirable and necessary for the largest scale manifestations, such as for
example the Bastille day flare), but by comparing global statistical
properties.

When searching for the answer to the first question, one must clearly use an
event definition which conserves the total energy in the signal, as one is
searching for a quantitative confirmation (again, much care is needed, since
the average corona exists to some extent precisely because we are uncapable
of observing fluctuations at sufficiently small energy and time-scale, \ie
it is by definition a background).  In the second case however, where there
are undoubtedly large differences between numerical models and observations
in the richness of the physics and dynamical range, one must be careful to
analyze events in the way most appropriate to glean characteristic
properties of the fluctuations and turbulence at the available scales.
Hence the requirements are to go beyond simply the energy distributions of
events and analyze other characteristic features such as anisotropy of the
spectra, intermittency, and higher order structure functions of the fields.

\begin{acknowledgements}
  The authors acknowledge partial financial support from the PNST (Programme
  National Soleil--Terre) program of INSU (CNRS) and from European
  Union grant HPRN-CT-2001-00310 (TOSTISP network). E.~Buchlin thanks the
  French-Italian University for travel support. The authors thank
  Jean-Claude Vial, Loukas Vlahos and Jean-François Hochedez for useful
  discussions.
\end{acknowledgements}

\bibliographystyle{natbib}
\bibliography{solphys,needtoget}

\end{document}